# The Recognizability of Individual Creative Styles Within and Across Domains

Liane Gabora
Brian P. O'Connor
Apara Ranjan

University of British Columbia

Corresponding Author:
Dr. Liane Gabora
Department of Psychology
University of British Columbia
Okanagan campus, Arts Building, 3333 University Way
Kelowna BC, V1V 1V7, CANADA
Email: liane.gabora@ubc.ca
Phone: 250-807-9849

**Abstract**

We present a set of studies that tested the hypothesis that creative style is recognizable within and across domains. Art students were shown two sets of paintings, the first by five famous artists and the second by their art student peers. For both sets, they guessed the creators of the works at above-chance levels. In a similar study, creative writing students guessed at above-chance levels which passages were written by which of five famous writers, and which passages were written by which of their writing student peers. When creative writing students were asked to produce works of art, they guessed at above-chance levels which of their peers produced which artwork. Finally, art students who were familiar with each other's paintings guessed at above-chance levels which of their peers produced which non-painting artwork. The findings support the hypothesis that creative styles are recognizable not just within but also across domains. We suggest this is because all of an individual's creative outputs are expressions of a particular underlying uniquely structured self-organizing internal model of the world.

**The Recognizability of Individual Creative Styles**
**Within and Across Domains**

This paper investigates two hypotheses concerning creative style. The first hypothesis is that creative style is a genuine construct, and that an individual's creative style is recognizable by others. Elsewhere the term *creative style* has been used to refer to an approach to problem solving and other tasks that is creative (e.g., Helson, 1967; Kumar, Kemmler & Holman, 1997; Mudd, 1995; Gelade, 2002; Goldsmith, 1987; Kirton, 1976; Runco & Basadur, 1993). Perhaps the most widespread use of the term is with reference to the Kirton Adaptation-Innovation Inventory, which measures the extent to which people generate ideas that violate accepted orthodoxy. Runco and Basadur (1993) identified different creative problem solving styles such as generator, conceptualizer, and optimizer. The focus in the present paper is not on kinds or dimensions of creative style, nor on personality traits associated with a creative approach to tasks, nor on the artistic styles of works created during particular epochs (i.e. Romantic, Gothic, Renaissance etc.). What we are interested in and what we mean by *creative style* is the extent to which an individual's creative outputs exhibit an identifiable character. This paper does not address *how* creative style is expressed, nor what makes it recognizable; we imagine that one's choice of techniques (e.g. heavy brush strokes) and numerous other subtler factors are involved. Our focus on whether it *is* recognizable.

The recognizability of creative styles is a phenomenon that may seem obvious to artists themselves and to those who appreciate what they do, but the topic has been largely neglected by psychologists. It has been shown that collections of art by a particular artist possess specific recurring stylistic features (Berlyne, 1971). However, there is little empirical, academic literature on this form of individual differences, and none that we know of that addresses the extent to which recurring stylistic features instill in the works an individual style that is reliably recognizable by others.

Our second hypothesis investigated is that if an individual's creative style is recognizable in one domain, it should be recognizable, to a lesser degree, in one or more other domains. Several studies (reviewed in Baer, 2010) have addressed a related but different question: whether an individual who is talented in one creative domain is talented in other creative domains. The hypothesis that individuals exhibit not *talent* but *personal creative style* through multiple domains is derived from the view that creative style is imparted by a core cognitive / personality structure that uses creative domains as conduits for its expression. Thus, style is not expected to be unique to a given domain but to come through in other domains, including the micro-creative decisions involved in everyday life and interaction with others.

We begin by discussing the theoretical framework underlying our approach and the relevant background material from the literature on creativity, in particular the issue of domain specificity / generality. We then present the studies themselves.

## Theoretical Framework Underlying the Current Approach

To our knowledge, no existing theory of creativity in its current form predicts or accounts for the phenomenon of enduring and identifiable creative styles. Central to the view of creativity that inspired these studies is the notion of a worldview. For lack of a better term, the word *worldview* is used to refer to one's internal model of the world, as well as one's values, attitudes, predispositions, and habitual patterns of response and affect (Gabora, 2000a, 2008; Gabora & Aerts, 2009). Thus we use the term in a broadly encompassing way that incorporates ones' understanding of, perspective on, and feelings about one's reality. It is posited that creativity arises due to the *self-organizing, self-mending* nature of a worldview (Gabora, under revision). Through the creative process one both (1) assimilates experiences into a more cohesive worldview, and (2) *expresses* or manifests this worldview. Thus this view emphasizes the internal cognitive and emotional restructuring brought about by the creative process and its impact on the individual.

Since the raw experiences one has are to a degree unique, and there are individual differences in the manner in which individuals assimilate and express their experiences, a worldview is expected to develop a characteristic structure, and the creator's outputs are different expressions of that structure. Therefore *creative outputs are expected to be related to one another, and to potentially pave the way for one another*. Moreover everyday instances of micro or min-creativity are expected to give evidence of, perhaps to a lesser degree, the characteristic structure of one's worldview (Beghetto & Kaufman, 2007). Thus the view that creativity involves mending and redefining one's unique cognitive / personality structure leads to the hypothesis that creative styles not only exist but are recognizable, within a domain and across domains.

This view of creativity is consistent with evidence that the creative process often involves extensive interpretation and transformation of material in a manner that is guided by and reflects one's personal experience and environment (Runco, 1996). The transformation that occurs on canvas or on the written page is mirrored by a potentially therapeutic sense of personal transformation and self-discovery that occurs within (Cropley, 1990; Maslow, 1971; May, 1976; Rogers, 1961; Singer, 2010). Both eminent creators and laypeople often claim that they gain a clearer sense of themselves as unique individuals from engaging in creative activities. Although for highly constrained problems with well-defined answers (such as when the fitness function is single-peaked), training and experience lead people to become more similar, the exact opposite is often the case in the arts (Ackerman, 2007). Artists often find a style that feels as if it is 'theirs' only after prolonged periods of exploration with different media and established styles and art forms (Ericsson, 1996). Similarly, writers and musicians speak of transitioning from a stage in which they were merely imitating the styles of creators they admired to a stage in which they felt they had discovered their own authentic 'voice' (Feinstein, 2006). The developing self-knowledge and sense of uniqueness is likely at least partially responsible for the tendency for highly creative individuals to have a more well-developed sense of who they are than less creative individuals (Fleith, Renzulli, Westberg, 2002; Garailordobil & Berrueco, 2007; MacKinnon, 1962).

**Domain Specificity and Domain Generality**

The second hypothesis tested in this paper throws a new angle on a longstanding debate about whether creativity is domain-specific or domain-general. Support for domain-specificity is provided by studies using techniques such as Amabile's (1982, 1983, 1996) Consensual Assessment Technique, showing that the tendency for expertise or eminence with respect to one creative endeavor to be only rarely associated with expertise or eminence with respect to another creative endeavor (e.g. Baer, 1993, 1996; Runco, 1989; Ruscio, Whitney, & Amabile, 1998; Tardif & Sternberg, 1988). For example, creative scientists rarely become famous artists or dancers.

On the other hand, as noted by Lubart and Guignard (2004), studies involving self-report scales, creativity checklists, and other sorts of psychometric or personality data tend to support the view that creativity is domain-general (Hocevar, 1976; Kaufman, Cole, & Baer, 2009; Runco, 1987; Plucker, 1998). The relevance of the latter studies to the general versus specific debate has been questioned because they do not actually measure creative outputs, but rather traits associated with the generation of creative output (Kaufman, Plucker, & Baer, 2008). However, cognitive reorganization and personality dynamics (e.g., involving wellbeing, self-discipline, or self-discovery) can be viewed as the *internal,* less readily measurable but equally important counterpart to the *external* manifestations of the creative process.

The apparent lack of genius-level versatility may be due in part to the “ten year rule” (Hayes, 1989), which is the finding that even highly talented creative individuals require approximately a decade of preparation before they are capable of generating a creative product that brings them to eminence. However, the above-mentioned evidence for domain-specificity is not limited to eminent creators but includes studies of typical research participants. In both cases, though, the focus is on *talent,* which is treated as unidimensional, as opposed to *style,* which is multidimensional. The underlying question is ‘are individuals talented in multiple domains?’ rather than ‘can individuals use multiple domains to meaningfully develop and express themselves?’.

Batey and Furnham’s (2006) comprehensive review of models and approaches at the intersection of personality, creativity, and intelligence studies provides a valuable synthetic framework for the integration of domain-general and domain-specific aspects of creativity. Theories that integrate domain-general and domain-specific aspects include Amabile’s (1983, 1996) componential model, Kaufman and Baer’s (2004) hierarchical model, and Eysenck’s (1993) biologically inspired model. Eysenck distinguishes between *originality,* a dispositional trait or cognitive style, which he claims is general in its application, and *exceptional achievement* in a creative domain, which he claims is highly specific in its application. He posits that the dispositional trait of originality is one of many factors that are necessary for exceptional creative achievement. We find the distinction is useful, at least as a first approximation, but the emphasis on the socially constructed yardstick of achievement presents problems. Many artists achieve success by generating variations on a successful theme instead of taking the risk of expressing themselves in an authentic and personally meaningful way, while others who do creatively express themselves in authentic and personally meaningful ways go unrecognized for some time. Undoubtedly there are creative geniuses who never achieved fame or recognition of any kind, who may not even have shared their creative work with anyone. Clearly, if someone is recognized posthumously, they do not switch from being creative in an original, domain-general way to being creative in a successful, domain-specific way. We view originality as a trait of the person or product, and achievement as a measure of the social recognition of originality, skill, or both. Although social recognition of achievement may be unlikely for the same individual in multiple domains, we suspect that individuals are capable of fulfilling self-exploration and expression in multiple creative domains, and that stylistic elements of these explorations are transported from one domain to another.

**The Present Studies**

The first two studies described below were conducted to provide evidence for the recognizability of creative styles within single domains of creativity: visual art (Study 1) and writing (Study 2). The focus was on the common belief that there *is* such a thing as recognizable creative style or voice. The third study tested the more stringent prediction that creative style is manifested in recognizable ways through *different* creative outlets.

## Study 1:The Within-Domain Recognizability of Artistic Styles

The first study tested the hypothesis that individuals who are familiar with the art (i.e., paintings or visual artworks) of a given artist will recognize other works by that artist that they have not encountered before.

### Method

**Participants.** The participants were three groups of fine arts majors consisting of 10, 14, and 15 students respectively, for a total of 39 students. The members of each group had been taking a studio art class together for at least one month. All students claimed to be highly familiar with five well-known artists. They also claimed to be highly familiar with the art of the other students within their group.

**Materials and procedure**. The participants were presented with two sets of paintings. For the first set, they were first shown three well-known paintings by each of five well-known artists as a refresher: Picasso, Monet, Van Gogh, Dali, and Andy Warhol. These artists were chosen because previous discussion with the students indicated that they were all familiar with these artists, but they each had a body of work that was large enough that we could find works the students would be highly unlikely to have encountered. The teachers also confirmed that the students had learned about the artists in class. So as not to bias the students' perceptions of the artists' styles with our own observations about their styles, there was no discussion with the students at any point about the particular styles of these artists. No paintings by artists other than these five were shown (i.e. no 'fillers'). The students were then shown ten unfamiliar works from the five artists – two by each artist – that they had not studied in class. The participants were told that there might be several paintings by one artist and none by another; each painting had an equal chance of being done by each artist. There was no discussion with the students either before or during the study about the particular styles of these artists.

For the second set, the participants were shown paintings by their fellow students that they had never seen before. Again, they were not told that there would be one painting by each classmate; each painting had an equal chance of being done by each classmate. The data for the three groups were kept separate for all of the analyses, including the analyses of the famous artists data. This was done for two reasons: (1) it should be less confusing to readers than would be a mixture of collapsed and non collapse-able findings for the groups; and (2) the findings for separate groups can provide evidence for the replicability of the data patterns across groups of respondents. (This practice was also used in Study 2 and Study 3.)

The signatures on all of the artworks were covered by a black bar. The participants were given a questionnaire that asked them to guess which famous artists were responsible for the paintings in the first set, and which classmates were responsible for the paintings in the second set. For each answer, they were also asked to state whether it was possible that they had encountered the work before. If this was the case, their data was discarded from the analyses. The percentage of guesses that were discarded for this reason ranged between 18% and 36% for the three groups.

**Analytic Methods**

For each set of paintings (i.e., for paintings by famous artists and for paintings by fellow students), two hit rate statistics were computed for each participant. The first was the simple hit rate, *H*, which is the proportion of correct guesses. Although intuitively meaningful, simple hit rates should be corrected for chance guessing and for response bias, such as the tendency to use particular response categories more or less than other response categories. The unbiased hit rate, *Hu*, corrects for both problems (Wagner, 1993). The *Hu* statistic is based on analytic methods from signal detection theory and involves the creation of a confusion matrix for each respondent, from which the *Hu* statistic is derived. For the present studies, we report both the mean simple hit rates (*H*) and the mean unbiased hit rates (*Hu*).

A data randomization procedure (Edgington, 1995; Good, 2005; Manly, 2007) was then used to assess the statistical significance of the mean *H* and *Hu* values. Specifically, for each set of paintings, each participant's guesses were randomly rearranged 1000 times and the *H* and *Hu* values were computed for each random data trial. The significance level was the proportion of trials that yielded *H* (or *Hu*) values greater than or equal to the *H* (or *Hu*) values for the real, non-randomized data. The mean *H* and *Hu* values for the random data sets were also used as the *H* and *Hu* values that would be expected on the basis of chance. These values served as the reference points for regular, parametric one-sample *t*-tests. Although there were slight differences in the *p* values for the data randomization and *t*-test procedures, the pattern of significant and non-significant findings was identical for the two kinds of significance tests for all three studies described in this article. The *r* correlation coefficient equivalent effect size for each *t* value was computed following Rosenthal and Rosnow (1991).

Multilevel modeling was used to assess the mean *H* and *Hu* values for the pooled data from the three groups. As described above, the 39 participants were divided into groups of 10, 14, and 15 for the second set of paintings (by classmates). When individuals are nested within groups in this way, their responses are potentially interdependent, and adjustments must be made for the statistical interdependence whenever data from different groups are pooled (de Leeuw & Meijer, 2008). An intercept-only mixed-effects model, using the nlme package in R, was used to obtain the adjusted mean *H* and *Hu* values (i.e., the intercepts and their significance tests) for the pooled data.

**Results**

**Recognition of works by famous artists.** The results for the within-domain recognizability of artistic styles are presented in Table 1. For the first group, the mean hit rates regarding which famous artist was responsible for which painting were $H = .79$ and $Hu = .75$. The mean hit rates that would have been obtained on the basis of random guesses for these questions were .20 and .14, respectively. Both hit rates were statistically significant, and the *r* effect sizes were large, .98 and .98. For the second group, the mean hit rates were high, $H = .76$ and $Hu = .68$, statistically significant, and the effect sizes were large, $r = .97$ and $r = .95$. For the third group, the mean hit rates were high, $H = .74$ and $Hu = .64$, statistically significant, and the effect sizes were large, $r = .98$ and $r = .95$. The pooled data hit rates were $H = .76$ and $Hu = .69$, both of which were statistically significant. The art students were thus able to indicate, at above-chance levels, which famous artists created which pieces of art that they had not seen before.

Table 1

Study 1: Mean Hit Rates, *t*-Test values, and *r* Effect Sizes for the Within-Domain Recognizability of Artistic Styles

| | **Mean Hit Rate** | **Chance Hit Rate** | ***t (df)*** | ***r* Effect Size** |
|---|---|---|---|---|
| **Art Students Recognition of Artworks By Famous Artists:** | | | | |
| Group 1 Hit Rate (*H*) | .79 | .20 | 13.1 (9) | .98 |
| Group 1 Unbiased Hit Rate (*Hu*) | .75 | .14 | 14.3 (9) | .98 |
| Group 2 Hit Rate (*H*) | .76 | .20 | 15.2 (13) | .97 |
| Group 2 Unbiased Hit Rate (*Hu*) | .68 | .16 | 10.7 (13) | .95 |
| Group 3 Hit Rate (*H*) | .74 | .20 | 18.5 (14) | .98 |
| Group 3 Unbiased Hit Rate (*Hu*) | .64 | .12 | 12.4 (14) | .96 |
| Pooled Data Hit Rate (*H*) | .76 | .20 | 27.4 (35) | .98 |
| Pooled Data Unbiased Hit Rate (*Hu*) | .69 | .14 | 18.1 (35) | .95 |
| | | | | |
| **Art Students Recognition of Artworks By Classmates:** | | | | |
| Group 1 Hit Rate (*H*) | .71 | .10 | 6.5 (9) | .91 |
| Group 1 Unbiased Hit Rate (*Hu*) | .80 | .17 | 8.2 (9) | .94 |
| Group 2 Hit Rate (*H*) | .60 | .17 | 5.0 (13) | .81 |
| Group 2 Unbiased Hit Rate (*Hu*) | .65 | .21 | 4.9 (13) | .81 |
| Group 3 Hit Rate (*H*) | .33 | .07 | 6.2 (14) | .86 |
| Group 3 Unbiased Hit Rate (*Hu*) | .55 | .11 | 5.6 (14) | .83 |
| Pooled Data Hit Rate (*H*) | .55 | .11 | 3.8 (35) | .54 |
| Pooled Data Unbiased Hit Rate (*Hu*) | .66 | .16 | 7.2 (35) | .77 |

Note. $p < .05$ for all mean hit rates and *t* values.

**Recognition of artworks by classmates**. For the first group, the mean hit rates regarding which classmate was responsible for which painting were $H = .71$ and $Hu = .80$, both of which were statistically significant with large effect sizes, $r = .91$ and $r = .94$. Similar findings emerged for the second group (mean $H = .60$, mean $Hu = .65$, with $r = .81$ and $r = .81$) and for the third group (mean $H = .33$, mean $Hu = .55$, with $r = .86$ and $r = .83$). The pooled data hit rates were $H = .55$ and $Hu = .66$, both of which were statistically significant. The art students thus also identified their classmates' art at above-chance levels.

## Study 2: The Within-Domain Recognizability of a Writer's "Voice"

The second study tested the hypothesis that individuals who are familiar with the works of particular writers will recognize other works by those writers that they have not encountered before.

### Method

**Participants.** The participants were 29 creative writing students who were familiar with five well-known writers and with each other's writing.

**Materials and procedures.** The procedure for the writers study was similar to the procedure described above for Study 1. Two sets of writings were presented to the participants. For the first set, they were first given three well-known written passages by each of ten well-known writers as a refresher. The well-known writers were Ernest Hemingway, Douglas Coupland, Emily Dickinson, Walt Whitman, Allen Ginsburg, Jack Kerouac, TS Eliot, Jane Austen, George Orwell, and Franz Kafka. These writers were chosen because previous discussion with the participants indicated that all students were familiar with these writers. We chose passages that were characteristic of the author's style without containing any obviously identifying information. Passages with proper nouns were included only if the proper noun was not associated with the author and not revealing of the author's identity (such as the name Amy). To illustrate, the refresher passage for Emily Dickinson was as follows:

> "You ask of my companions. Hills, sir, and the sundown, and a dog large as myself, that my father bought me. They are better than beings because they know, but do not tell; and the noise in the pool at noon excels my piano. I have a brother and sister; my mother does not care for thought, and father, too busy with his briefs to notice what we do. He buys me many books, but begs me not to read them, because he fears they joggle the mind. They are religious, except me, and address an eclipse, every morning, whom they call their "Father." (Dickinson, p. 254)."

There was no discussion with the students either before or during the study about the particular styles of these writers. The participants were then shown twenty rare passages – two from each of the ten famous writers – that they had not studied in class. To illustrate, the rare passage for Ernest Hemmingway was as follows:

> "He was a very complicated man compounded of absolute courage, all the good human weaknesses and a strangely subtle and very critical understanding of people. He was completely dedicated to his family and his home and he loved more to live away from them." (Hemmingway, 1999, p. 1).

The participants were told that there might be two by one writer and none by another; each passage had an equal chance of being written by each writer. Passages were chosen in such a way as to avoid proper nouns that would be associated with a particular writer.

For the second set, the participants were shown a piece of writing by each of their fellow classmates that they had never seen before. Again, they were told that each piece of writing had an equal chance of being done by each classmate. In order to make sure that participants were familiar with each others' work, the 29 participants were divided into groups of eight, seven, six, and eight for the second set of writings. We requested that students avoid proper nouns and any identifying information in their passages. The participants were given a questionnaire that asked them to guess which famous writers were responsible for the writings in the first set, and which classmates were responsible for the writings in the second set. For each answer, they were also asked to state whether they might have encountered the passage before. Whenever this was the case, their data were discarded from the analyses. The percentage of guesses that were discarded for this reason ranged between 2% and 14% for the four groups. The analytic methods for Study 2 were the same as those used for Study 1.

## Results

**Recognition of writings by famous writers**. The results for the within-domain recognizability of a writer's "voice" are presented in Table 2. Similar results were obtained for the four groups of participants in their attempts to identify passages of text written by famous writers. They are easily summarized. The simple hit rate values, *H*, ranged between .31 and .41. The unbiased hit rates, *Hu*, ranged between .22 and .43. All of the hit rates were well above their corresponding chance hit rates and were statistically significant. The *r* effect sizes were all large, ranging between .82 and .94. The pooled data hit rates were *H* = .37 and *Hu* = .29, both of which were statistically significant. The creative writing students thus identified, at above-chance levels, the authors of passages of text written by famous writers that they had not encountered before.

Table 2

Mean Hit Rates, *t*-Test values, and *r* Effect Sizes for the Within-Domain Recognizability of a Writer's "Voice" for Study 2.

| | **Mean Hit Rate** | **Chance Hit Rate** | ***t (df)*** | ***r* Effect Size** |
|---|---|---|---|---|
| **Writing Students Recognition of Text By Famous Writers:** | | | | |
| Group 1 Hit Rate (*H*) | .31 | .10 | 7.4 (7) | .94 |
| Group 1 Unbiased Hit Rate (*Hu*) | .22 | .06 | 4.9 (7) | .88 |
| Group 2 Hit Rate (*H*) | .34 | .10 | 4.8 (6) | .89 |
| Group 2 Unbiased Hit Rate (*Hu*) | .22 | .06 | 3.9 (6) | .85 |
| Group 3 Hit Rate (*H*) | .40 | .10 | 4.7 (5) | .90 |
| Group 3 Unbiased Hit Rate (*Hu*) | .28 | .05 | 3.7 (5) | .86 |
| Group 4 Hit Rate (*H*) | .41 | .10 | 4.2 (7) | .85 |
| Group 4 Unbiased Hit Rate (*Hu*) | .43 | .11 | 3.8 (7) | .82 |
| Pooled Data Hit Rate (*H*) | .37 | .10 | 9.5 (24) | .89 |
| Pooled Data Unbiased Hit Rate (*Hu*) | .29 | .07 | 4.2 (24) | .65 |
| | | | | |
| **Writing Students Recognition of Text By Classmates:** | | | | |
| Group 1 Hit Rate (*H*) | .32 | .13 | 4.0 (7) | .83 |
| Group 1 Unbiased Hit Rate (*Hu*) | .23 | .07 | 4.4 (7) | .86 |
| Group 2 Hit Rate (*H*) | .33 | .14 | 3.0 (6) | .78 |
| Group 2 Unbiased Hit Rate (*Hu*) | .37 | .17 | 3.2 (6) | .79 |
| Group 3 Hit Rate (*H*) | .67 | .34 | 2.2 (5) | .70 |
| Group 3 Unbiased Hit Rate (*Hu*) | .67 | .36 | 2.0 (5) | .67 |
| Group 4 Hit Rate (*H*) | .34 | .13 | 1.7 (7) | .55 |
| Group 4 Unbiased Hit Rate (*Hu*) | .41 | .19 | 1.4 (7) | .47 |
| Pooled Data Hit Rate (*H*) | .41 | .17 | 2.9 (24) | .51 |
| Pooled Data Unbiased Hit Rate (*Hu*) | .41 | .19 | 2.6 (18) | .47 |

Note. $p < .05$ for all mean hit rates and *t* values.

**Recognition of writings by classmates.** Similar results were obtained for passages written by the four groups of the students themselves. The simple hit rate values, *H*, ranged between .32 and .67. The unbiased hit rates, *Hu*, ranged between .23 and .67. All of the hit rates were well above their corresponding chance hit rates and were statistically significant, and the *r* effect sizes were substantial, ranging between .47 and .86. The pooled data hit rates were $H = .41$ and $Hu = .41$, both of which were statistically significant. Thus the creative writing students also identified, at above-chance levels, passages of text written by their classmates that they had not read before.

## Study 3: The Recognizability Of Creative Styles Across Domains

Our third study tested the hypothesis that familiarity with an individual's creative work in one domain aids the recognition of that individual's creative work in another domain.

### Method

**Participants**. There were six sets of participants in Study 3. The first four sets (Ns = 7, 7, 6, & 8) consisted of undergraduate students in different sections of an advanced creative writing class. The two remaining sets of participants (Ns = 13 & 15) consisted of undergraduates majoring in art.

**Materials and procedure.** The creative writing students were asked to submit one or more piece of covered, unsigned art prior to the study. They were explicitly asked not to submit art with anything that would obviously identify them as the artist. (For instance, in one class, following the suggestion of the professor of that class, the example was given that a particular member of the class who was obsessed with surfing should not give any indication of that obsession.) The creative writing students were familiar with each others' writing, but not with each others artwork. The artworks were brought to a subsequent meeting of the class and the students were shown the unsigned art done by their classmates.

The art students were asked to submit one or more non-painting artwork prior to the study. The art students were familiar with each others' paintings, but not with each others' non-painting artworks. The non-painting artworks they submitted included poetry, a scarf, a clay pot, a jute wall-hanging, a photograph, and a stuffed toy.

The professor brought all of the artworks to class and the students were shown the unsigned art done by their classmates. The participants were given a questionnaire and asked to guess which classmate did which piece of art. For each answer, they were also asked to state whether it was possible that they had encountered the work before. Whenever this was the case their data was discarded from the analyses. The percentages of guesses that were discarded for this reason ranged between zero% and 19% for the six groups. The analytic methods for Study 3 were the same as those used for Study 1.

### Results

**Creative writing students' recognition of artworks by classmates.** The results for the recognizability of creative styles across domains are presented in Table 3. There was some variation in the findings across the four groups, with the results for Group 1 and Group 3 being very similar and statistically significant, which was in contrast with positive but non significant results for Group 2 and Group 4. Specifically, for Groups 1 and 3, the *H* and *Hu* values ranged between .59 and .74, they were statistically significant at the .05 level, and the *r* effect sizes ranged between .76 and .81. For Groups 2 and 4, the hit rates were notably lower (*H* = .22 and *Hu* = .24 for Group 2; and *H* = .19 and *Hu* = .25 for Group 4), non significant at the .05 level, and the effect sizes were weaker (.33 and .20 for Group 2, and .38 and .33 for Group 4). However, the pooled data hit rates, *H* = .43 and *Hu* = .48, were both statistically significant. The findings thus provide mixed evidence at the group level, but clear evidence at the pooled data level, that creative writing students can identify, at above-chance levels, which of their classmates created a given work in a domain other than writing.

Table 3

Study 3: Mean Hit Rates, *t*-Test values, and *r* Effect Sizes for the Recognizability Of Creative Styles Across Domains

| | **Mean Hit Rate** | **Chance Hit Rate** | ***t (df)*** | ***r* Effect Size** |
|---|---|---|---|---|
| **Writing Students Recognition of Art By Classmates:** | | | | |
| Group 1 Hit Rate (*H*) | .59 | .17 | 3.2 (6) | .80 |
| Group 1 Unbiased Hit Rate (*Hu*) | .70 | .22 | 3.4 (6) | .81 |
| Group 2 Hit Rate (*H*) | .22 | .16 | 0.9* (6) | .33 |
| Group 2 Unbiased Hit Rate (*Hu*) | .24 | .20 | 0.5* (6) | .20 |
| Group 3 Hit Rate (*H*) | .72 | .34 | 2.9 (5) | .79 |
| Group 3 Unbiased Hit Rate (*Hu*) | .74 | .39 | 2.6 (5) | .76 |
| Group 4 Hit Rate (*H*) | .19 | .13 | 1.1* (7) | .38 |
| Group 4 Unbiased Hit Rate (*Hu*) | .25 | .19 | 0.9* (7) | .33 |
| Pooled Data Hit Rate (*H*) | .43 | .19 | 1.8 (23) | .35 |
| Pooled Data Unbiased Hit Rate (*Hu*) | .48 | .24 | 1.8 (23) | .34 |
| | | | | |
| **Art Students Recognition of Non-Painting Works By Classmates:** | | | | |
| Group 1 Hit Rate (*H*) | .53 | .16 | 3.5 (12) | .71 |
| Group 1 Unbiased Hit Rate (*Hu*) | .64 | .21 | 4.0 (12) | .77 |
| Group 2 Hit Rate (*H*) | .36 | .08 | 4.6 (14) | .78 |
| Group 2 Unbiased Hit Rate (*Hu*) | .73 | .19 | 5.8 (14) | .84 |
| Pooled Data Hit Rate (*H*) | .44 | .12 | 3.9 (25) | .61 |
| Pooled Data Unbiased Hit Rate (*Hu*) | .69 | .20 | 6.9 (25) | .82 |

Note. $p < .05$ for all mean hit rates and *t* values except for *, which indicates $p > .05$.

**Arts students' recognition of non-painting artworks**. For Group 1, the mean hit rates were *H* = .53 and *Hu* = .64, and the hit rates that would have been obtained on the basis of random guesses were *H* = .16 and *Hu* = .21. Both mean hit rates were statistically significant, and the *r* effect sizes were large, .71 and .77. Similarly, for Group 2 the hit rates were .36 and .73, both of which were statistically significant, and the effect sizes were .78 and .84. The pooled data hit rates were *H* = .44 and *Hu* = .69, both of which were statistically significant. Thus the art students were thus able to identify, at above-chance levels, which of their classmates created a given work in a domain other than painting.

## General Discussion

The results supported the hypothesis that creative individuals have a distinctive style or 'voice' that is recognizable both within and across creative domains. Art students were able to identify at above-chance levels which famous artists created which pieces of art they had not seen before. They also identified their classmates' art at above-chance levels. Similarly, creative writing students were able to identify, at above-chance levels, passages of text written by famous writers that they had not encountered before, and passages of text written by their classmates that they had not encountered before. Furthermore, creative writing students were generally able to identify works of art produced by classmates that they had not seen before, and art students who were familiar with only the paintings of other art students were able to identify non-painting artworks by the other students at above-chance levels.

The ability of creative writing students to identify works of art produced by classmates that they had not seen before occurred at a statistically significant level for two of our four groups of participants. The mean hit rates for the remaining two groups were in the correct direction but they were two-to-three times lower than the hit rates for the other two groups. We suspect that the weaker effects in two of the groups were due to the presence of a small number of persons who performed extraordinarily poorly. The mean hit rate for the pooled data was nevertheless significant, indicating an overall effect. The lack of statistical significance for two of the groups should be considered in its proper context. Small groups were required to ensure that the students were familiar with each others' work, but small Ns reduce statistical power. The non-significant effect sizes that emerged for our small groups were nevertheless on par with effect sizes typically reported in psychological research. It is also highly unusual for researchers to report findings for small sub-groups of their data sets, as we have done. Researchers typically report findings only for their pooled data, and the pooled-data findings in the present studies were all statistically significant.

We also found that the effect sizes for the within-domain recognizability of artistic styles (Study 1) were notably stronger than the effect sizes for the within-domain recognizability of a writer's "voice" (Study 2). A possible reason for the difference in effect sizes is that more information about a creator's style is available in artworks than in short passages of text.

Although it may be possible to account for these results with existing theoretical frameworks for creativity, the results are not predicted by other theories, and it is not straightforward how they would be accommodated by other theories, particularly those that emphasize chance processes or the accumulation of expertise. Such theories provide no reason to expect that the works of a particular creator should exhibit a unique and recognizable style. Nor do they account for findings reported elsewhere that the act of creation leads to an enhanced sense of self (Fleith et. al., 2002; Garailordobil & Berrueco, 2007; MacKinnon, 1962). These findings are, however, predicted by the view that creativity is the process by which a worldview is forged and expressed, according to which personal style reflects the uniquely honed structure of an individual's worldview. The fact that creative writing students were able to identify which of their classmates created a given work in a domain *other* than writing, and that art students were able to identify the creators of non-painting artworks that they had not seen before, supports the prediction that individuals exhibit styles that are recognizable not just within a domain but also across domains, although still within what Baer and Kaufman (2005) identify as the same thematic cluster, i.e. the arts.

A related finding is that artistically naïve individuals can perceive cross-media stylistic similarities, and group together instances of art by different individuals done in the same genre or from the same temporally defined period, e.g. Neoclassical (Hasenfus, Martindale & Birnbaum, 1983). Of perhaps even greater interest with respect to the results of Study Three is Peretti's (1972) finding that music students were able to match paintings by Paul Klee with the musical selections that inspired Klee to paint them. Since Klee did not compose the music himself, this study does not address cross-domain *individual* style. It does, however, speak to the complexity of the phenomenon, and specifically the possibility that style may be subject to many subtle influences including exposure to works in seemingly unrelated domains.

The evidence reported here that creative artifacts are manifestations of an underlying personal style highlight the importance of a view of creativity that balances internal and external effects of the creative process. The two are expected to be related, and there is evidence that they are. For example, self-report measures of the ability to experience original, appropriate combinations of emotions (referred to as emotional creativity) is correlated with laboratory and self-reported creativity measures (Ivcevic, Brackett, & Mayer, 2007). However, focusing on the product and neglecting the internal change brought about by the creative process can lead to a distorted view of creativity. For example, it is commonly assumed that the creative individual is compelled "to transform the creative idea into a creative product" (Thrash, Maruskin, Cassidy, Fruer & Ryan, 2010, p. 470). This may be the case, but to the intrinsically motivated creator, the product may function primarily as a way of tracking the progress of the idea; its worldly value may be of secondary importance. Indeed while the tradition in the West is to focus almost exclusively on creativity as the process by which a new and useful or entertaining product is generated, Eastern conceptions focus more on creativity as a process that can bring about therapeutic change (Niu & Sternberg, 2001), i.e. that expresses, transforms, solidifies, or unifies the creator's understanding of and/or relationship to the world. In the extreme, the external creative work can be viewed as a mere byproduct of the internal transformation brought about through engagement in a creative task.

A theory of creativity that focuses on external results of the creative process cannot explain common attitudes toward creative artifacts, such as that while an original masterpiece is viewed as creative, a reproduction or imitation of the masterpiece is generally not. A view of creativity that balances internal and external aspects can make sense of this. Only the original masterpiece provided humanity with a newfound roadmap to understanding or expressing something. Once the focus shifts from the products resulting from the creative process to its transformative effects, the success of a creative venture cannot be measured in purely objective terms. A creative venture may be successful though it does not generate something of external value if it facilitates the discovery and expression of the individual's potentially unique and thus recognizable patterns of thought or emotion.

The present studies also provide a new perspective on the controversy over whether creativity is domain-general or domain-specific. It is widely assumed that the question of whether creativity is domain-specific or domain-general can be resolved by determining to what extent ratings of expertise in one domain are correlated with ratings of expertise in another. Measurements of expertise are assumed to be sufficient to detect any quality that might characterize or unify an individual's creative ventures, and creative outputs are assumed to be objectively comparable. This study differs from previous research that addresses the domain-specificity / generality question in that the focus has been not on talent but on personal creative style. Creative achievement can be characterized in terms of not just expertise or eminence but the ability to express what we genuinely are through whatever media we have at our disposal at a given time. One might expect that an artist's or scientist's personal style will be evident in how he or she prepares a meal, decorates a room, or expresses a personal experience, what creativity researchers refer to as little-c creative activities (Gardner, 1993).

The present studies suggest that it is appropriate to incorporate into our understanding of the interplay between domain-specific and domain-general mechanisms an underlying unified yet multi-faceted personality structure and way of assimilating and responding to the world, which characterizes an individual's creative output. The expertise that is required for creative success makes it likely that creativity will appear domain-specific when the focus is on the usefulness or entertaining features of creative products. But creative products should nevertheless have a style that is recognizable within and across creative domains that reflects the worldviews of their creators. Higher cognition may be domain-general, not in the sense that expertise in one enterprise guarantees expertise in another, but in the sense that there are multiple interacting venues for creative expression open to an individual, and through which that individual's worldview may be gleaned.

The support obtained here for the hypothesis that domain generality manifests as not (necessarily) talent in multiple domains, but as the expression of personal style through multiple domains, is of social significance, for it has potential implications for both education and our understanding of the factors involved in wellbeing. If individuals have multiple interacting avenues open to them for creative self-expression, and if the creative process is of value to the extent that it not just (1) yields products deemed by experts to be of high quality, but (2) exerts a positive, transformative effect on the creator, opportunities for domain general learning are potentially of significant social consequence. It may be that our capacity for cross-domain learning is just beginning to be exploited, through ventures such as the Learning through the Arts program in Canada, in which students, for example, learn mathematics through dance, or learn about food chains through the creation of visual art. If knowledge is *presented* in compartmentalized chunks, then students may end up with a compartmentalized understanding of the world. If knowledge were presented more holistically, a more integrated kind of understanding may be possible.

The creative styles phenomenon also provides a relatively novel and challenging form of individual differences for personality researchers. Most dimensions of individual differences are introduced with descriptions of the relevant characteristics of persons at the high and low ends of a latent trait continuum. A trait is usually expected to have a near-normal distribution, and a measure of the trait is typically provided to stimulate research. But creative styles and worldviews are like multidimensional fingerprints. They are characterized by unique blends or constellations of features. Single measures and normal distributions are not likely or relevant. The creative styles phenomenon is idiographic and perhaps very difficult to measure with self-report questionnaires. The present research nevertheless indicates that creative styles are clearly recognizable from creative outputs, both within and across creative domains.

## Limitations and Future Research

There are several limitations of this study, some of which are unavoidable and some of which we hope to address in future studies. An unavoidable limitation is the small number of subjects in each study. In each case the number of subjects was determined by the number of students who showed up for class on the day of the study. It was necessary to use small classes of students rather than an ad-hoc group gathered for the purpose of the study to ensure familiarity with each other's work. The larger the class, the less familiarity they would have with each other's work. We do not think this presents a problem for interpreting the data given the large effect sizes, the replications of the effects across the sub groups of participants in each study, and the consistently significant effects for the pooled data.

Another limitation is that the examples of famous artists and writers come from different eras, and thus in part the students were likely discriminating amongst conventions associated with a particular era as opposed to individual creative style *per se*. The examples were chosen following consultation with both the professor of the class and the students themselves as to which artists or writers' work they were familiar with and believed they would actually recognize. Given that the classes were studio classes and creative writing classes, without a component involving study of the works of others, it was more difficult than anticipated to find a set of artists or writers that they unanimously agreed were familiar and recognizable. To find sets from the same era would simply not have been possible. It is possible that this could be overcome by conducting these experiments at a university that requires all students to take particular art history or literature courses. At any rate, we believe that the conventions associated with a particular era become woven into, and become part of, an individual's creative style, and thus that the use of examples from different eras does not invalidate the findings reported here. While we asked that passages written by students not contain proper nouns, and we avoided passages by famous authors containing proper nouns that could aid in the identification of the author, in future studies we will avoid proper nouns in the written passages entirely.

A third limitation is that the reporting of whether a creative work had been encountered before was dependent on self-report. It is possible that participants had encountered works by the famous creators before and forgotten about them. This criticism, however, is much less likely to apply to the works created by the participants' classmates, for which significant results were also obtained.

Could recognizability be due to differences in familiarity with, and use of, various techniques, as opposed to differences in creative style? We do not think so. With respect to the studies involving recognition of works by student peers, the students were in the same class, and indeed in the same program, and were thus exposed to the same techniques. They nonetheless produced very different products. The concern is more applicable to the studies involving recognition of works by eminent creators, who may well have been taught different techniques. However, even for these studies, we believe it is highly unlikely that participants were only recognizing differences in technique. The issue is analogous to arguments about form versus content. It seems reasonable that one gravitates toward techniques, tools, instruments, and so forth that provide opportunities to express what one wants to express. Over time, the things one wants to express and the tools with which one expresses them become intertwined. In any case, an important topic for future research is to clarify the relationship between creative style and technique.

A more general limitation is that this kind of result is only expected in domains where there is room for expression of individual or cultural differences in personalities and worldviews. To the extent that a creative domain or task is highly constrained (in complexity theory terms, to the extent that the fitness function is single-peaked) one would not expect distinctly recognizable styles to emerge.

In future work we plan to conduct a similar study using eminent artists from the same time period and from within a single art school or tradition. This would enable us to assess to what extent individual creative style is influenced by the environment in which they create. Another direction for future studies is to use a similar approach to investigate the recognizability of creative styles in (a) individuals who have developed proficiency in multiple creative domains, and (b) individuals who for disability or health reasons become unable to continue working in their original creative domain and subsequently switch to a second creative domain. If their creative styles are still recognizable in the second domain, this will provide further support for the hypothesis that creative style reflects an underlying personality structure.

## Acknowledgments

This work was funded by grants to the first and second authors from the Social Sciences and Humanities Research Council of Canada (SSHRC) and a grant to the first author from the Concerted Research Program of the *Flemish Government of Belgium*. We would like to thank Shawn Serfas, Sharon Thesen, Anne Fleming, Michael Smith, Gary Pearson, and particularly Nancy Holmes, as well as their students, for their assistance in carrying out this research. Finally, we thank the editor and anonymous reviewers for their astute comments and suggestions.